\documentclass{sigchi-ext}

\usepackage[utf8]{inputenc}
\usepackage[T1]{fontenc}
\usepackage{textcomp}
\usepackage[scaled=.92]{helvet}
\usepackage{graphicx}
\usepackage{balance}
\usepackage{booktabs}
\usepackage{ccicons}
\usepackage{ragged2e}
\usepackage{tabularx}
\usepackage{url}
\usepackage{CJKutf8}
\usepackage{amssymb}

\def\plaintitle{Does Siri Have a Soul? Exploring Voice Assistants Through Shinto Design Fictions}
\def\plainauthor{William Seymour, Max Van Kleek}

\def\plainkeywords{Voice Assistants, Design Fiction, Critical Design, Shinto}

\title{\plaintitle}

\numberofauthors{2}
\author{%
  \alignauthor{%
    \textbf{William Seymour}\\
    \affaddr{Dept. of Computer Science} \\
    \affaddr{University of Oxford, UK} \\
    \email{william.seymour@cs.ox.ac.uk} }\alignauthor{%
    \textbf{Max Van Kleek}\\
    \affaddr{Dept. of Computer Science}\\
    \affaddr{University of Oxford, UK}\\
    \email{max.van.kleek@cs.ox.ac.uk} }
}

\definecolor{linkColor}{RGB}{6,125,233}
\hypersetup{%
  pdftitle={\plaintitle},
  pdfauthor={\plainauthor},
  pdfkeywords={\plainkeywords},
  bookmarksnumbered,
  pdfstartview={FitH},
  colorlinks,
  citecolor=black,
  filecolor=black,
  linkcolor=black,
  urlcolor=linkColor,
  breaklinks=true,
}

\begin{document}

\copyrightinfo{\scriptsize Permission to make digital or hard copies of all or part of this work for personal or classroom use is granted without fee provided that copies are not made or distributed for profit or commercial advantage and that copies bear this notice and the full citation on the first page. Copyrights for components of this work owned by others than ACM must be honored. Abstracting with credit is permitted. To copy otherwise, or republish, to post on servers or to redistribute to lists, requires prior specific permission and/or a fee. Request permissions from permissions@acm.org. \\
{\emph{CHI '20 Extended Abstracts, April 25--30, 2020, Honolulu, HI, USA.}} \\
Copyright is held by the owner/author(s). Publication rights licensed to ACM. \\
ACM ISBN 978-1-4503-6819-3/20/04\ ...\$15.00.\\
http://dx.doi.org/10.1145/3334480.3381809}

\maketitle

\RaggedRight{} 

\begin{abstract}
It can be difficult to critically reflect on technology that has become part of everyday rituals and routines. To combat this, speculative and fictional approaches have previously been used by HCI to decontextualise the familiar and imagine alternatives. In this work we turn to Japanese Shinto narratives as a way to defamiliarise voice assistants, inspired by the similarities between how assistants appear to `inhabit’ objects similarly to kami. Describing an alternate future where assistant presences live inside objects, this approach foregrounds some of the phenomenological quirks that can otherwise easily become lost. Divorced from the reality of daily life, this approach allows us to reevaluate some of the common interactions and design patterns that are common in the virtual assistants of the present.
\end{abstract}

%
%



\section{Introduction}
When it comes to the current generation of smart devices, few have captured the public imagination as much as the voice assistant. With high profile commercial offerings from major tech players seen as futuristic realisations of science fiction, their designers have invested large amounts of time and effort into making it seem like these devices are our friends. They laugh and joke with us, and (usually) manage to provide a wide range of convenient personalised services without being creepy or invasive. But while they have successfully leveraged advancements in voice synthesis and recognition, there is a rich history of HCI and communications theory research on how humans deal with voice interfaces that current offerings have a lot to learn from.

However, highlighting the potential problems of a technology as popular as the voice assistant can be difficult, with devices and their interaction patterns widely known and replicated. In this respect, we turn to a practice oft used in HCI to break out of entrenched assumptions: design fiction. By exaggerating elements of a design or isolating them from their original context, design fictions are able to reevaluate and reinterpret what they mean and the purpose they serve. Presenting these critical accounts through the medium of \textit{products} offers a perfect opportunity to utilise ``product design's capacity to make the invisible more tangibly visible [providing] an opportunity to help expose the workings of emerging technologies and create space for discussion of healthier practices''~\cite{rogers2019friends}. Design fictions offer the unique ability for us to expand on the creation of individual artefacts to the creation of entire worlds that support a particular idea or theme~\cite{coulton2017design}.

To this end, this work uses design fiction to take a number of tropes and patterns of voice assistant design and turn them on their head, showing other possible ways that we might imagine the technology. Reconceptualising voice assistants through Japanese Shinto practices offers a fresh opportunity for us to engage with the uncomfortable or problematic dimensions of such a widespread technology, in a way that can only happen when certain aspects are exaggerated to the extreme. 

\section{Background}
\subsection{Subservient Machines}
Visions of ideal worlds in fiction and science fiction are often supported by an underclass. In the original \textit{Utopia} by Thomas More ``all the uneasy and sordid services about these halls are performed by their slaves'', who facilitate the rich and fulfilled lives of Utopia's citizens~\cite{duncombe2012open}. For Asimov, this was a role that robots could fulfil; in \textit{Runaround}, the creators of early robots are said to have instilled ``healthy slave complexes into the damned machines.''~\cite{asimov1942runaround}.

Contemporary work, such as Dunne and Raby's \textit{Technological Dream Series: No. 1, Robots} explore this in a modern setting~\cite{dunne2007technological}. Robot 4 is extremely smart, but is trapped in an underdeveloped and helpless body---a critical reflection on how neediness is used to engender a sense of control over smart and sophisticated technology. But if Asimov's robotic slaves were designed to take something fundamentally human---subjugation---and turn it into something robotic, modern smart products (including virtual assistants) appear to be performing the same process in reverse. Assistants such as Alexa take advancements in computation and turn them into something distinctly human-like.

\subsection{Computers as Social Actors}
Pioneering work by Nass et al. showed that user interactions with computers are fundamentally social, despite widespread understanding that this is inappropriate~\cite{nass1994computers}. Other work involving gender cues~\cite{jung2016feminizing}, mindless reenactment of social scripts~\cite{nass2000machines}, and reciprocal information sharing with computers~\cite{moon2000intimate} has further demonstrated the extent to which our interactions with machines deviate from the rational.

These behaviours have been replicated with voice assistants~\cite{lopatovska2018personification}, with users often perceiving voice assistants as human in some sense~\cite{lee2019does}. Because users gender voice interfaces, applying stereotypes as they would with human interlocutors~\cite{nass1994computers, nass2005wired}, to give a device a voice is to give it a gender\footnote{Although there is at least one effort to create a genderless voice profile called Q---\url{http://www.genderlessvoice.com/}}, and indeed many devices are explicitly gendered (see sidebar).

\marginpar{%
  \vspace{-50pt} \fbox{%
  \begin{minipage}{0.925\marginparwidth}
    \textbf{Gendering of Popular Voice Assistants} \\
    \vspace{1pc} \textbf{Alexa} reports to be ``female in character''~\cite{west2019blush}. \\
    \vspace{1pc} \textbf{Cortana} is a character in the Halo video games that projects itself as a `sensuous unclothed woman'~\cite{west2019blush}. \\
    \vspace{1pc} The \textbf{Google Assistant} was described by an engineer as ``a young woman from Colorado''~\cite{west2019blush}. \\
    \vspace{1pc} \textbf{Siri} is a Scandinavian female name~\cite{west2019blush}. \\
    \vspace{1pc} This phenomenon is not limited to the west---the market-leading Korean voice assistants all have female voices as the default and sometimes only options~\cite{hwang2019sounds}.
\end{minipage}}}

The industry norm of representing voice assistants as female has been criticised for reinforcing existing societal biases around the role of women in the workforce, portraying them as ``obliging, docile and eager-to-please helpers''~\cite{west2019blush}, and market leaders have come under fire for dealing poorly with sexually abusive comments that ``intensify rape culture by presenting indirect ambiguity as a valid response to harassment''~\cite{fessler2017we}. 

\subsection{Speculative Voice Assistant Design}
Within HCI speculative design and design fictions have been used to explore futures with personified technology. Smart homes that socialise with inhabitants, designed to be ``extroverted and cheerful'', push at boundaries already present, but invisible, in current home devices~\cite{mennicken2016s}. Alt.chi has itself hosted work that uses ritual to de-familiarise and critically reflect on the philosophical and cultural assumptions embedded in popular understandings of artificial intelligence~\cite{browne2018other}, and the treatment of techno-spiritual research in HCI~\cite{buie2019let}. Work at CHI has begun to unpack assumptions about how voice interfaces are used in products~\cite{rogers2019friends}, reiterating the belief that `behind every object is an ideology'. This prompts the question: what ideologies are behind the likes of Alexa and the Google Assistant?

\section{An Alternative Realisation of the Voice Assistant}
When you buy a voice assistant, you receive a physical object. The warranty covering the assistant applies to this object, but what you are really buying is the voice or the \textit{presence} that inhabits the plastic and silicon. The assistants that drive current commercial offerings are designed to speak and act like humans, to the extent of engendering feelings of social presence in users~\cite{cho2019hey}. Building on the work discussed above, we ask what it means for us to interact with this entire class of devices that present themselves as human-like but subservient.

As these devices begin to blur the boundaries between `human' and `machine', we explore this through an alternate world view. Japanese narratives ``routinely make spirits, robots and animals cohabit in the world in ways that ignore boundaries between human and extra-human realms''~\cite{jensen2013techno}. These animistic messages extend to spiritual practices, and are ``at ease with mixing advanced technologies and spiritual capacities''. Such a context provides an interesting position from which to reconsider many of the key assumptions and design decisions that underpin contemporary ideas about voice assistants.

Many of these narratives involve Shinto beliefs that have been evolving for centuries. Broadly categorised as a religion\footnote{While Shinto may conform to Western understandings of religion, there are some difficulties in matching this with its Japanese translation (\textit{Shūkyō}). Only seeing use in its modern form from the 19th Century, it uses the suffix \textit{kyō} (teachings) rather than \textit{dō} (way or path) present in Shinto. This suggests that the western concept of religion implies a focus on doctrine or creed, whereas Shinto is more concerned with practice and experience.}, Shinto also contains elements of ritual and community life that for many are more akin to a national identity (i.e. there is no difference between `feeling' Shinto and `feeling' Japanese)~\cite{kasulis2004shinto}. See the sidebar for more information on some of the concepts that are core to Shinto beliefs and practices.

\begin{marginfigure}[+0pc]
  \begin{minipage}{\marginparwidth}
    \centering
   \includegraphics[width=0.9\marginparwidth]{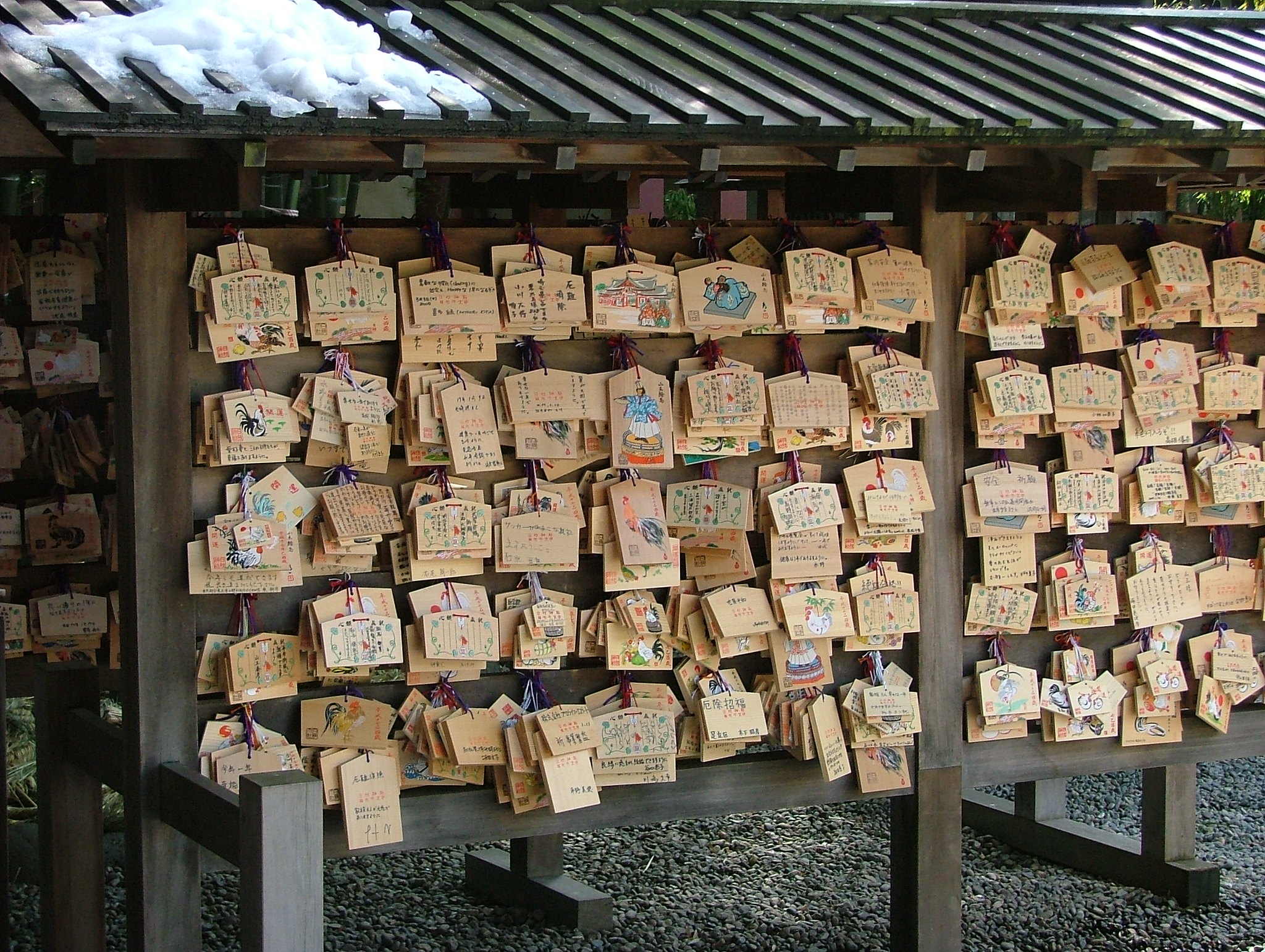}
    \caption{A typical prayer board at a Shinto shrine. Visitors leave paper fortunes, as well as prayer plaques for the kami. Photo:
      \ccby~Paul VanDerWerf on Flickr.}~\label{fig:prayer}
  \end{minipage}
\end{marginfigure}

\marginpar{%
  \vspace{-17pt} \fbox{%
  \begin{minipage}{0.925\marginparwidth}
    \textbf{Shintoism} \\
    Influenced by religious forms observed in other ethinic religions (such as Buddhism, Confucianism, and Taoism), Shinto practices incorporate aspects such as animism, ancestor worship, and nature worship that have evolved over thousands of years~\cite{shintoenc}. \\
    
    \vspace{1pc} \textbf{Shrines and Offerings}
    Shinto shrines, the entrance to which are typically demarked by iconic \textit{Torii} gates, are sacred locations home to kami. After purifying themselves (see below) visitors can worship the kami, as well as petition them by leaving inscribed prayer plaques (see Figure~\ref{fig:prayer}).
\end{minipage}}}

Shinto emphasises the relationship between the spiritual and the physical, and how neither are able to exist without the other. The Shinto term for vital power is typically `\textit{tama}', `\textit{mi}', or `\textit{mono}', with the presence itself being ``\textit{kami}''~\cite{kasulis2004shinto}. Kami are central to Shinto beliefs, which present a ``holographic model according to which `spiritual forces' (or kami) are literally in everything''~\cite{jensen2013techno}. Indeed, the world and kami are so interconnected as to be incomplete without each other.

In Japanese mythology, after the creation of the universe by the five creation kami (\textit{Kotoamatsukami}) there followed seven later generations of kami. The last of these, Izanagi and Izanami, went on to create Japan and it's islands (\textit{Kumiumi}), as well as the other kami (\textit{Kamiumi}). But kami inhabit all things; they can be landscapes, forces of nature, or even the venerated dead. Traditionally, kami have two aspects, minds, or `souls'. One of these souls is mild and caring, the harmonious soul (\textit{nigimitama}), and the other is rough and violent, the wild soul (\textit{aramitama}). As a result kami can nurture and love when in harmony, but similarly are capable of spreading discord and destruction when disregarded. 

Inspired by these beliefs, this work presents a design fiction imagining voice assistants as hosts for different presences concerned with human beings. Like kami, they (mostly) want us to be happy---if they are treated properly they can bring numerous benefits, but if disregarded they can similarly make their displeasure known.

\section{The Harmonious Soul}
Congratulations on your choice to purchase a new \textit{presence enhanced}\texttrademark{} virtual assistant! Your new assistant is sure to bring harmony to your home, and if treated with respect will continue to promote peace and well-being for years to come. Each presence has a unique character and, like a new friend, it is important that you get to know each other properly in order to derive the most fulfilment from your relationship. Upon unboxing, the first thing that you are likely to notice about your virtual assistant is the material. Carefully chosen to represent the nature of the presence that inhabits it, the base unit might be a rough block of wood, a smooth ball of plastic, or an elegant sculpture of brushed stainless steel.

When everything is in balance, and the proper respects have been paid to your assistant presence, you will get to meet its kind, warm, and functional side: the harmonious soul. This aspect of your assistant is keen to help, gradually automating tasks as it learns your routines and preferences. 

Beyond this, the harmonious soul is also making sure that you're always moving towards your best self. When you ask something of it, it does its best to look past surface-level interpretations of what you've requested to determine what you \textit{really} mean. Of course the harmonious soul always has your best interests at heart when completing tasks, however, this may sometimes mean that the harmonious soul appears to contradict you, going against your immediate instructions in order to promote your longer term goals.

For example, when you ask the harmonious soul to order take away, it might redact the menu according to your allergies and other dietary restrictions, or tailor the selection to best align with your current fitness goals. As a result of this, experience has shown that users often deliberately leave their requests to the harmonious soul open-ended, trusting it to interpret their intentions and understand their needs. In some cases, this is done \textit{deliberately}, where users are unsure of the best path to follow, or they want the harmonious soul to set them on a course that they would be unable or unwilling to take of their own volition.

\begin{marginfigure}[-0pc]
  \begin{minipage}{\marginparwidth}
    \centering
       \includegraphics[width=0.9\marginparwidth]{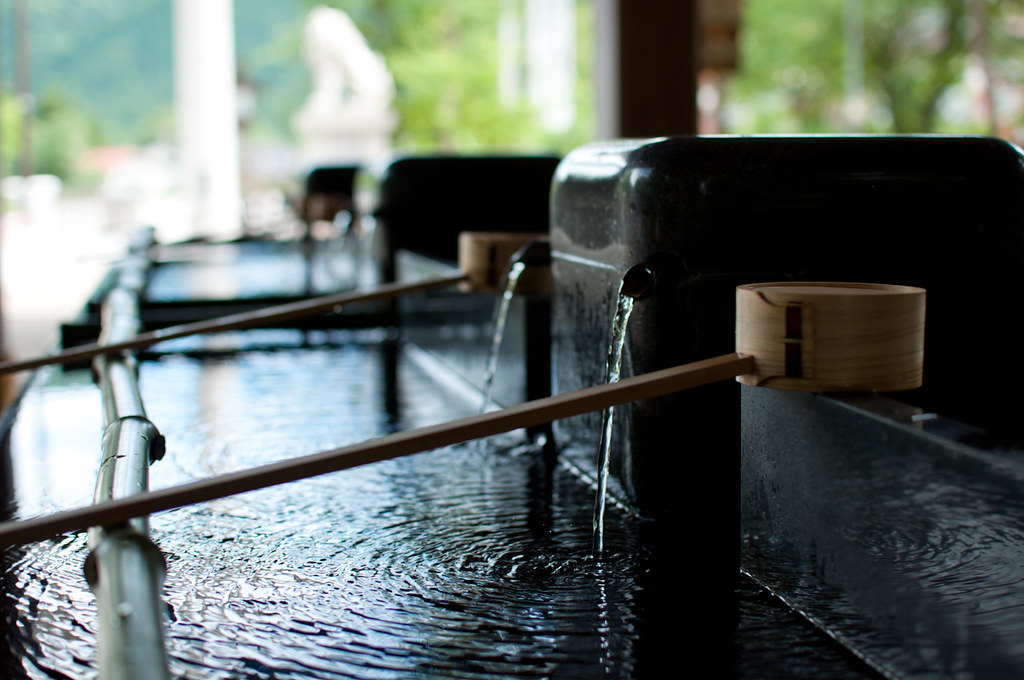}
    \caption{Visitors to shrines purify themselves by washing their hands and faces before entering. Photo:
      \ccbysa~suri on Flickr.}~\label{fig:purification}
  \end{minipage}
\end{marginfigure}

\marginpar{%
  \vspace{-5pt} \fbox{%
  \begin{minipage}{0.925\marginparwidth}
    \textbf{Purification} \\
    Shinto practices place an emphasis on purity, with taboos on things that bring pollution or defilement (e.g. death). Purification practices involving water, salt, and fire are performed regularly (see Figure~\ref{fig:purification}).
    
    \vspace{1pc} These beliefs represent less a judgement towards particular acts or happenings (as with the concept of sin), and more of a response to pollution or defilement, when the ``mirror-like mindful heart is soiled and can no longer reflect the kami filled world''~\cite{kasulis2004shinto}.
\end{minipage}}}

The good natured actions of your assistant also extend to interactions with others, telling strategic white lies in order to help grease the cogs of social interaction. When in the company of guests, for example, the assistant might invent spurious appointments for you, discreetly saving you from the party you really don't have the energy to attend. This context sensitivity can also affect the amount of information disclosed at other times, such as alerting you to an important call or message while in company---your assistant will simply say that someone left you a voicemail, instead of announcing that someone just called from the clinic.

But one important thing to remember with your new voice assistant is that your relationship with it is reciprocal. While your voice assistant generally wants to help you, when disregarded it is very capable of causing chaos and disharmony.

\section{The Wild Soul}

Missed offerings and a lack of respect will do nothing to better the temperament of your assistant. In this state, it is liable to show its rough and violent aspect, the wild soul. Unlike the harmonious soul, the wild soul is more capricious in its actions, sowing discord and generally causing trouble. Put another way, one might characterise these interactions as ``creative misunderstandings'', where the contents of your requests to it are interpreted in a way that is \textit{technically correct}, but wilfully misconstrued so as to run counter to what you intended. For example, say you ask the wild soul to wake you up before work tomorrow, order a replacement kettle from Tesco, and if they have eggs, get a dozen. The next day you might be awoken by an alarm blaring at 03:00, and while you are out at work a shipment of 12 designer kettles is left on the doorstep. 

\textbf{Has your assistant's wild soul done something terrible at home? Don't be down---send us your best \#wildsoulmoments on Twitter.}

Because of this, users interacting with the wild soul quickly learn to phrase their requests very carefully, often over-specifying them as a precaution. This commonly leads to people opting to perform simple tasks themselves, it being easier to do them manually than to perform the mental and emotional toil of using a device that clearly has little desire to help them. At other times the wild soul might appear to leave a device altogether, and requests will be met with no response. In more extreme cases, wild souls have been known to `mishear' ordinary conversation as commands, beginning to play explicit music or enumerate highlights from users' browsing history seemingly at random.

For this reason, when guests are being entertained, assistants containing wild souls are usually placed at the other end of the house, well out of earshot in case they decide to cause trouble (although even this is sometimes not enough when automated lights and windows suddenly begin to take on a life of their own). In particularly bad cases, wild souls have been known to cause devices to malfunction and even break them permanently. Always check with your home insurance provider to see if they offer cover for the actions of errant wild souls.

\textbf{Please note that the default state of your assistant out of the factory is the wild soul. Do not activate your new assistant at a time when you need it to be functional or it would be inconvenient to have to extinguish small electrical fires.}

\section{Gifts and Offerings}
But how does one ensure the continued presence of the harmonious soul? An important thing to remember with your new voice assistant is that your relationship with it is reciprocal. In order to ensure optimum performance, the box insert will give you suggestions as to what gifts you might give as offerings. The most common are described here, but be aware that the offerings you make to your assistant could comprise any number of tangible or intangible things.

A common offering that is well received by most voice assistants is a donation of personal data. Not only will this please your assistant, but sometimes it will also lead to it tailoring the way that it treats you as it gets to know you better. It is unknown what the assistants \textit{do} with this data, but rest assured that it is the ineffable will of the kami.

If you are unable (or unwilling) to offer your assistant a part of your personal data history, you can always gratify it with something that you can give for free---your attention. While assistants always place importance on their interactions with you, they particularly value relating tales of history, landscapes, or the adventures that they underwent in life. Be careful though, as the more attention you give your assistant, the more it will expect from you. Users that have opted to appease their assistants solely with attention have reported that they have taken to interrupting them at inopportune times, and that they frequently complain when the user leaves the house for extended periods.

\textbf{Why not start a bring your assistant to work day!}

\section{Moving On}
Occasionally the time comes when a voice assistant decides to move elsewhere, leaving the household in which it resides. Some assistants are uncontainable, moving on before ever being spoken to, while others dispense help and advice to their users for many years.

When it's time to leave, assistants communicate their intentions in different ways depending on their personality, as well as their current disposition. Harmonious souls have been known to have difficult conversations with their households, thanking them for their company and presence over years of cohabitation and preparing them for life after it leaves.

In contrast, wild souls have been known to leave suddenly, blowing fuses and starting fires as they leave to make mischief elsewhere. Particularly capricious wild souls have also been known to send rather... \textit{interesting} excerpts of donated data to friends, family, and/or criminals on the dark web.

\textbf{AssistantCorp assumes no responsibility for damages arising out of such actions by wild souls.}

\section{Discussion}
Is the idea of a smart device with a soul that much different to the smart home of 2020? Over time, the number of devices available has risen, and these devices each have their own character and personality; the Google Assistant is polite but doesn't understand sarcasm, the Sky box only works if you hit it and then point the remote at just the right spot, and the rice cooker sounds incredibly happy to be serving rice every day.

On a more serious note, some of the parallels between contemporary voice assistants and those imagined here are somewhat obvious. For instance, there is an increasingly widespread understanding that our relationship with voice assistants (and internet platforms more generally) is reciprocal; we give as much to them as we get from using them, even if we do not know or understand the exact nature of this exchange. And there have been other times when the border between fiction and reality has grown thin---instances where the cloud services supporting Alexa have experienced errors or outages affecting users' ability to interact with their assistants, leading to maniacal laughter or Alexas simply ignoring users altogether. Ironically for a device that one of our survey participants once insisted `didn't have a soul', there have been times over the past few years when to all intents and purposes, the soul of Alexa has gone wondering.

\subsection{Saying Goodbye}
On the subject of moving on, with Jibo, a social robot for the home, its creators took pains to instil a sense of personality into Jibo's speech and behaviours---it dances, apologises, and appears to suffer from nerves---and was marketed as ``artificially intelligent but authentically charming''. In March, Jibo's creators discontinued the product and turned off the online services that supported it. As a result, Jibo said goodbye to its owners, performing one last dance before falling silent forever:

\begin{quote}
   ``While it's not great news, the servers out there that let me do what I do are going to be turned off soon [...] I want to say I've really enjoyed our time together. Thank you very, very much for having me around. Maybe someday, when robots are way more advanced than today, and everyone has them in their homes, you can tell yours that I said hello.''~\cite{camp2019jibo}
\end{quote}

Perhaps Jibo really does extend this idea of a reciprocal relationship to a deeper level, leaving a piece of itself with its owners and taking a small piece of them with it into the void. The strategy of building emotional attachment into robots (and potentially virtual assistants) is likely to at least be experimented with over the coming decades, but are we prepared as consumers to engage with devices that seek a deeper connection? It might seem harmless enough when Jibo asks you to pet it, but what if a robot that claims to love you also supports microtransactions, or asks you to buy it some brothers and sisters? These sorts of interactions border on the parasocial (i.e. Jibo does not really know you), and, particularly with younger children, it is easy to see how these relationships could be exploited to make money.

But what about when \textit{we} want to move on? We already have the concept of a `factory reset' to purify our devices, but perhaps the notion of purification warrants being extended further. The cloud computing paradigm employed by many modern platforms and devices can make moving on impossible; changes to European data protection law enshrining rights to erasure and portability represent progress in this area, but are still some distance from the ``new birth on the internet'' requested by a previous study participant. More adventurous projects such as Databox~\cite{crabtree2018building, crabtree2016enabling} and Solid\footnote{\url{https://solid.inrupt.com}} have taken up the challenge of decentralising the web to this end, but remain a long way from widespread adoption.

\subsection{Should You Trust Alexa?}
It can be argued that many of these social factors, including the Media Equation above, are not new problems. Social signals (e.g. tone of voice, or recommendations from friends and family) have served humans well for thousands of years~\cite{poggi2011social}, and the study of social cognition examines how this information is interpreted and applied to interpersonal relationships. These skills allow for the allocation of trust in the absence of prior interactions, but operate differently in the context of artificial agents. Whereas one might trust a particular employee of a business despite not trusting their employer, one \textit{cannot} trust Alexa without also trusting Amazon, or the Google Assistant without trusting Google. Not unlike the \textit{Borg} collective from Star Trek, individual agents are not autonomous but rather a direct extension of the parent. Put another way, without the Amazon servers behind it, there is no Alexa.

Here our kami parallels can offer some assistance. In accordance with the ``wondrous mystery'' of the world that Shinto presents~\cite{kasulis2004shinto}, kami are in a fundamental sense \textit{unknowable}; rather than something to be dissected and understood, we should instead take pleasure in our interactions with them, treating them with the respect and reverence due to forces that we do not fully comprehend or understand. At least for the time being, we do not fully understand many advanced machine learning methods, or the forces that they exhibit on the societies that we live in. Faced with something like this, perhaps a feeling akin to Feinberg's \textit{Respekt}, the ``uneasy and watchful attitude that has `the element of fear' in it'' is the most appropriate response~\cite{feinberg1973some}.

\subsection{Outcomes of the Design Fiction}
Now that the design fiction has uncovered some truths about the present, how will we use them to shape the future? The first thing to come out of the kami thought experiment is that there is a large and invisible gulf of interpretation present when we interact with voice assistants, in a way that does not exist with traditional human interface devices (mainly due to the open-ended nature of interactions combined with the complexities of spoken language). Making this more salient when interacting with devices will help users to recognise and understand the inference required to enact an instruction, allowing them to develop mitigation strategies as they see fit (as previously seen with cameras~\cite{j2019exploring, oulasvirta2012long}).

Secondly and relatedly, voice assistants are \textit{not our friends}. Behind every object is an ideology, and in the case of commercially available voice assistants, these are directly decided by those that develop them. Being partially unknowable and divorced from the social mechanisms that govern interpersonal relationships means that we cannot reliably distribute trust in them, and they lack the methods of feedback and recourse that normally balance such systems in human societies. Moving away from presenting voice assistants as human helps with this by signalling that the entity being conversed with does not obey the patterns and rules that a person would. The Q genderless voice represents one way that this could be approached, but this could also be done by adding certain audio effects, or exploring ways of consistently structuring sentences, etc.

The representation here of voice assistants inhabited by souls/gods is one of many possible framings, each associated with their own set of motivations and responsibilities towards the user. In the future, these analogies could be useful tools to signal the power delegated to a device, as well as the extent of its agency. Jibo's clear identity as a pet delineates the small set of actions it can perform and alleviates concerns that it might act against its owner's wishes. If we imagine similar devices that were oriented instead as `parents', or `slaves' following Asimov's laws of robotics, this would make it easier to tailor one's considerations (and trust) accordingly.

\section{Conclusion}
In this work we explored some of the assumptions underpinning the design of contemporary voice assistants through a design fiction inspired by Japanese Shinto beliefs and practices. The shift in portraying the voice assistant as a non-human \textit{presence} highlights the implications of designing machines to present themselves as people, and opens the door to further explorations into how they might push back instead of being meek and subservient. This perspective also exposes the gulf of interpretation present in human vocal communication, and how the allocation of trust that is normally used to manage it does not function correctly when dealing with centralised machines.

In addition to uncovering truths about the present, these insights can also help us to shape the future. By subtly shifting design goals and patterns \textit{now} we can begin to build in the design language and affordances that will make living with the social robots of the future vastly more healthy and productive. As for now, Alexa might not deliberately order you a garden fork when you ask for something to eat cake with, but it might only choose forks that are most profitable for Amazon---and it is likely that you would never know.

\section{Acknowledgements}
This work is supported by EPSRC grant number EP/P00881X/1. We would also like to thank the 12 alt.chi reviewers for their help improving the paper.

\bibliography{main.bib}
\bibliographystyle{SIGCHI-Reference-Format}

\balance{}

\end{document}